\documentclass[journal=jacsat,manuscript=article]{achemso}

\usepackage[version=3]{mhchem} 

\author{Bruno Ipaves}
\affiliation{Center of Natural and Human Sciences, Federal University of ABC (UFABC), Santo André, 09280-560, São Paulo, Brazil}
\email{ipaves.bruno@ufabc.edu.br}
\author{João F. Justo}
\affiliation{Escola Politécnica, University of São Paulo (USP), São Paulo, 05508-010, São Paulo, Brazil}
\author{James M. de Almeida}
\affiliation{Ilum School of Science, Brazilian Center for Research in Energy and Materials (CNPEM), Campinas, 13083-970, São Paulo, Brazil}
\author{Lucy V. C. Assali}
\affiliation{Institute of Physics, University of São Paulo (USP), São Paulo, 05508-090, São Paulo, Brazi}
\author{Pedro Alves da Silva Autreto}
\affiliation
{Center of Natural and Human Sciences, Federal University of ABC (UFABC), Santo André, 09280-560, São Paulo, Brazil}
\email{pedro.autreto@ufabc.edu.br}

\title[An \textsf{achemso} demo]
  {Enhancing catalyst activity of two-dimensional C$_4$N$_2$ through doping for the hydrogen evolution reaction}

\abbreviations{IR,NMR,UV}

\begin{document}

\begin{abstract}
This study investigates the structural, electronic, and catalytic properties of pristine and doped C$_4$N$_2$ nanosheets as potential electrocatalysts for the hydrogen evolution reaction. The pristine C$_{36}$N$_{18}$ nanosheets exhibit limited HER activity, primarily due to high positive Gibbs free energies ($>$ 2.2 eV). To enhance catalytic performance, doping with B, Si, or P at the nitrogen site was explored. Among these systems, B-doped C$_{36}$N$_{17}$ nanosheets exhibit the most promising catalytic activity, with a Gibbs free energy close to zero ($\approx$ -0.2 eV), indicating efficient hydrogen adsorption. Band structure, projected density of states, charge density, and Bader charge analyses reveal significant changes in the electronic environment due to doping. While stacking configurations (AA$'$A$''$ and ABC) have minimal effect on catalytic performance, doping — particularly with B —substantially alters the electronic structure, optimizing hydrogen adsorption and facilitating efficient HER. These findings suggest that B-doped  C$_{36}$N$_{17}$ nanosheets could serve as efficient cocatalysts when combined with metallic materials, offering a promising approach to enhance catalytic efficiency in electrocatalytic and photocatalytic applications.
\end{abstract}

\section{Introduction}

Given the increasing global energy demand and environmental crisis, developing sustainable energy systems to produce clean fuels and chemicals has gained increasing interest. These systems are essential to replace fossil fuels and mitigate carbon dioxide emissions, which are major contributors to climate change \cite{liao2022density, hasani2019two}. Among clean energy carriers, hydrogen (H$_2$) stands out due to a unique set of characteristics, such as non-toxicity, renewability, zero pollution, high abundance, and the highest energy density per unit mass, making it a promising candidate for a primary energy source \cite{chodvadiya2021enhancement, hasani2019two}. 

Nevertheless, the widespread adoption of hydrogen as a sustainable energy carrier depends on its production from clean and renewable sources, often referred to as "green hydrogen." Green hydrogen, generated through processes such as water electrolysis powered by renewable energy, has the potential to contribute significantly to the decarbonization of energy systems \cite{kourougianni2024comprehensive, hassan2024green}. In addition, the ability of hydrogen to store energy efficiently positions it as a key solution for large-scale energy storage, addressing the problem of intermittent renewable energy sources, such as solar and wind. This dual role — as both a clean fuel and a medium for energy storage at the grid level — underscores its importance in the transition toward carbon-neutral power systems \cite{kourougianni2024comprehensive, hassan2024green}.

In this context, developing efficient electrocatalysts for the hydrogen evolution reaction (HER) is critical to enable green hydrogen production through water electrolysis. Among the materials investigated, two-dimensional (2D) systems have garnered significant attention since the isolation of graphene, due to their unique properties \cite{sun2019high, hasani2019two}. These materials provide large surface areas, tunable electronic properties, and chemical versatility, making them ideal for several applications, including batteries \cite{ipaves2023aluminum}, sensors \cite{slathia2024ultralow, chakraborty2024subpicomolar}, oxygen reduction reaction (ORR) \cite{lucchetti2024cerium}, and HER \cite{sun2019high, yu2018c}. Consequently, the investigation of 2D materials and the exploration of novel strategies, such as doping or defect engineering, to enhance their catalytic properties, has become a key focus in the search for sustainable energy solutions.

Among 2D materials, carbon nitride-based nanosheets have shown considerable promise in enhancing HER efficiency by providing abundant active sites for hydrogen adsorption and improving reaction kinetics \cite{sun2019high, yu2018c, hasani2019two, chodvadiya2021enhancement}. Recently, C$_4$X$_2$ (X = B or N) nanosheets have been explored through first-principles calculations, revealing their intriguing structural and electronic properties \cite{ipaves2024tuning}. However, their potential catalytic activity for HER remains unexplored, leaving a gap in understanding their suitability as catalysts for sustainable hydrogen production.

In the present work, we investigate the structural, electronic, and catalytic properties of pristine C$_4$N$_2$ nanosheets using first-principles calculations, along with boron (B), silicon (Si), and phosphorus (P)-doped systems. Our study evaluates the Gibbs free energy of hydrogen adsorption, electronic structure modifications, and the doping role in optimizing the HER performance, providing insight into the potential of C$_4$N$_2$ nanosheets as efficient catalysts.

\section{Computational details}

This study used first-principles calculations based on spin-polarized Density Functional Theory (DFT) \cite{hohenberg1964inhomogeneous, kohn1965self}, utilizing the plane-wave basis set and projector augmented-wave (PAW) method \cite{kresse1999ultrasoft}, as implemented in the Quantum ESPRESSO software package \cite{giannozzi2009quantum,giannozzi2017advanced}. The generalized gradient approximation of Perdew–Burke–Ernzerhof (GGA-PBE) exchange-correlation functional \cite{perdew1996generalized} was applied, along with the Dion {\it et al.} scheme \cite{dion2004} refined by Klimeš {\it et al.} (optB88-vdW \cite{klimevs2009}) to accurately account for van der Waals (vdW) interactions. The plane-wave energy cutoff was set at 80 Ry, with a total energy convergence threshold of 0.1 meV/atom. To sample the irreducible Brillouin zone \cite{monkhorst1976special} a $6 \times 6 \times 1$ $k$-point mesh was utilized.

We constructed hexagonal supercells based on the primitive cell of the 2D structures previously investigated 
\cite{ipaves2024tuning}. The cell parameters in the $xy$-plane were determined through variable cell optimization using the BFGS quasi-Newton algorithm. To avoid interactions between cell images, we fixed the lattice parameter perpendicular to the sheets (along the $z$-axis) at 25 Å. This approach has proven effective in analogous 2D systems in prior studies \cite{garcia2011group, ipaves2019carbon, ipaves2022functionalized}. To investigate the interaction between the H atom and the doped C$_{36}$N$_{17}$, we initially placed H at approximately 1 {\AA} from the dopant.

We investigated the HER activities of pristine and doped nanosheets using the Sabatier principle \cite{chodvadiya2021enhancement}. According to this principle, the catalyst must bind strongly to the intermediate to facilitate the reaction. In contrast, the product (hydrogen atoms) should be weakly bound to the surface to allow for rapid desorption. At equilibrium, the efficiency of HER catalysis on a surface is dictated by the exchange current density, which is related to the Gibbs free energy $\Delta G$ under standard conditions (acidic medium, pH $\rightarrow$ 0), and can be defined as follows \cite{chodvadiya2021enhancement}:
\begin{equation}
    \Delta G = \Delta E_{\rm{ads}} + E_{\rm{ZPE}} - T \Delta S,
    \label{eq.gibbs}
\end{equation}
\noindent
where $\Delta E_{\rm{ads}}$ is the adsorption energy of a hydrogen (H) atom, $E_{\rm{ZPE}}$ is the zero-point energy difference between hydrogen atoms in the adsorbed and gas phases, which ranges from 0.01 to 0.04 eV, and $\Delta S$ is the entropy change, which is approximately 0.24 eV. The hydrogen adsorption energy is given by \cite{chodvadiya2021enhancement}
\begin{equation} 
\Delta E_{\rm{ads}} = E{\rm_{(System + H)}} - E_{\rm (System)} - \frac{1}{2} E_{({\rm H}_{2})}, 
\label{eq.Eads} 
\end{equation}
\noindent
where $E{\rm_{(System + H)}}$ is the total energy of hydrogen adsorbed on pristine or doped systems, $E_{\rm (System)}$ is the energy of pristine or doped systems without H adsorption, and $E_{({\rm H}_{2})}$ is the energy of an isolated hydrogen molecule. Therefore, Equation \ref{eq.gibbs} can be approximated by\cite{chodvadiya2021enhancement}
\begin{equation} 
\Delta G =\Delta E_{\rm{ads}} + 0.24 \ \text{eV}.
\label{eq.deltag} 
\end{equation}

The work function of our systems was determined  using the expression
\begin{equation} 
\Phi = V_{\infty} - E_F, 
\label{eq.EP}
\end{equation}
\noindent
where $\Phi$ is the work function, $E_F$ is the Fermi energy, and $V_{\infty}$ is the vacuum level or electrostatic potential. For semiconductor systems, $E_F = (E_{\rm VBM} + E_{\rm CBM})/2$, where VBM is the valence band maximum and CBM is the conduction band minimum. 

We computed the charge density differences for H adsorbed on pristine or doped nanosheets to better understand the charge transfer. These differences were determined using the equation
\begin{equation}
    \Delta \rho = \rho_{\rm{(System + H)}} - \rho_{\rm{(System)}} - \rho_{\rm{(H)}}
    \label{eq:charge_diff}
\end{equation}
\noindent
where $\rho_{\rm{(System + H)}}$, $\rho_{\rm{(System)}}$, and $\rho_{\rm{(H)}}$ are, respectively, the charge densities of H adsorbed on pristine or doped systems, pristine or doped systems, and isolated H atom. Bader charge analysis was also performed to quantify the charge transfer from the H atom to the C$_4$N$_2$ \cite{tang2009grid, yu2011accurate}.

\section{Results and discussion}

\begin{figure}[hbt]
    \centering
    \includegraphics[height=7.3cm]{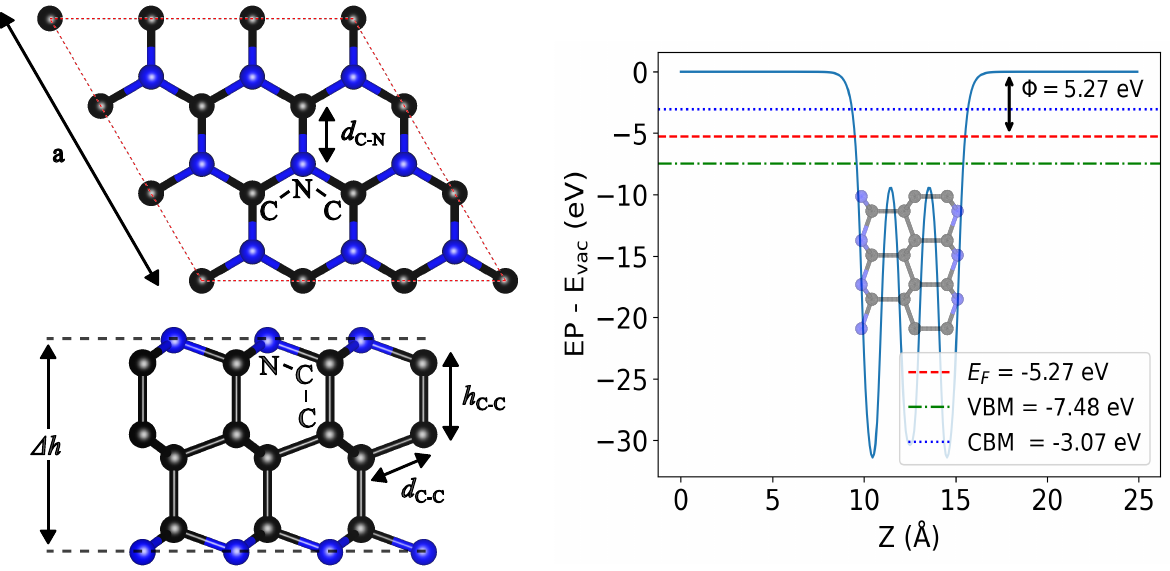}
    \caption{Schematic representation, presented in top and side views (left), and work function $\Phi$ (right) of optimized pristine AA$'$A$''$-C$_{36}$N$_{18}$ nanosheet. Structural parameters (lattice constants, bond lengths, and bond angles) are highlighted. The respective values are given in Table \ref{tb:1}. The work function, Fermi energy, VBM, and CBM are referenced relative to the vacuum energy level. The corresponding details for the ABC stacking configuration are provided in Figure S1 in the SI.}
    \label{fig:structure_pristine}
\end{figure}

We initially optimized a $3\times3\times1$ supercell of the C$_4$N$_2$ nanosheets, considering both AA$'$A$''$ and ABC stacking configurations. The supercell consisted of 54 atoms, including 36 carbon (C) and 18 nitrogen (N) atoms. Since both stacking arrangements exhibited similar structural and electronic properties, we present all figures for the ABC stacking in the Supporting Information (SI).  Figures \ref{fig:structure_pristine} and S1 in the SI  illustrate the configuration of these systems, while Table \ref{tb:1} provides the corresponding structural parameters. 

The lattice parameters are 7.27 and 8.66 Å for the AA$'$A$''$ and ABC configurations, respectively. For the AA$'$A$''$ stacking configuration, the intralayer C–C bond length ($d_{\rm C-C}$) and interlayer C–C distance ($h_{\rm C-C}$) are 1.49 and 1.60 Å, respectively. The thickness ($\Delta h$) is 4.74 Å, and the C–N bond length ($d_{\rm C-N}$) is 1.49 Å. In the ABC stacking configuration, the corresponding values are 1.50, 1.56, 4.66, and 1.49 Å, respectively. The C–C–N bond angle is 110.07° for the AA$'$A$''$ stacking configuration, while the C–N-C bond angle is 108.86°. These angles are, respectively, 109.84° and 109.10° for the ABC stacking configuration.

\begin{table*}[t]
\small
\centering
\caption{Structural and electronic properties of pristine and hydrogen-adsorbed C$_{36}$N$_{18}$: lattice parameter ($a$), intralayer ($d$) and interlayer ($h$) distances, thickness ($\Delta h$), intraplanar (C-N-C) and interlayer (C-C-N) bond angles, electronic band gap ($E_g$), work function ($\Phi$), and Gibbs free energies ($\Delta G$), as labeled in Figures \ref{fig:structure_pristine} and \ref{fig:structure_pristine_H} (AA$'$A$''$) and S1 and S2 in the SI (ABC). Distances are given in {\AA}, energies in eV, and angles in degrees.}
\begin{tabular}{lllllllllccc}
 System & \multicolumn{1}{c}{$a$} &\multicolumn{3}{c}{Bond length} & \multicolumn{1}{c}{$\Delta h$} &  H Distance & \multicolumn{2}{c}{Angle} & $E_g$ & $\Phi$ & $\Delta G$ \\
\hline
 & \multicolumn{1}{c}{} &\multicolumn{3}{c}{} & \multicolumn{1}{c}{} &  & \multicolumn{2}{c}{} & & &  \\ [-3mm]
  AA$'$A$''$     & &$d_{\rm C-C}$ & $h_{\rm C-C}$ & $d_{\rm C-N}$ &  &          & C-C-N & C-N-C &                 &                 &                 \\
& \multicolumn{1}{c}{} &\multicolumn{3}{c}{} & \multicolumn{1}{c}{} &  & \multicolumn{2}{c}{} & & &  \\ [-3mm]
Pristine & 7.27 & 1.49 & 1.60 & 1.49 & 4.74 &  & 110.07 & 108.86  & 4.41 & 5.27 &  \\
H C$_{\rm Top}$ & 7.27 & 1.49 & 1.60 & 1.49 & 4.74 & 3.23 (C-H) & 110.07 & 108.86  & 4.22 &  & 2.59 \\
H Hollow & 7.27 & 1.49 & 1.60 & 1.49 & 4.74 & 3.21 (R-H) & 110.07 & 108.86 & 4.21 &  & 2.59 \\ 
H N$_{\rm Top}$ & 7.32 & 1.50 & 1.60 & 2.05 & 4.83 & 1.03 (N-H) & 112.85 & 113.80  & 0.60 &  & 2.23 \\ \\
ABC  &   &  &  &  &    &         &  & &                 &                 &                 \\
 & \multicolumn{1}{c}{} &\multicolumn{3}{c}{} & \multicolumn{1}{c}{} &  & \multicolumn{2}{c}{} & & &  \\ [-3mm]
Pristine & 7.30 & 1.50 & 1.56 & 1.49 & 4.66 &  & 109.82 & 109.12  & 4.14 & 5.32 &  \\
H C$_{\rm Top}$ & 7.30 & 1.50 & 1.56 & 1.49 & 4.66 & 3.23 (C-H) & 109.82 & 109.12  & 3.96 &  & 2.59 \\
H Hollow & 7.30 & 1.50 & 1.56 & 1.49 & 4.66 & 3.20 (R-H) & 109.82 & 109.12  & 3.97 &  & 2.59 \\ 
H N$_{\rm Top}$ & 7.36 & 1.50 & 1.56 & 2.05 & 4.79 & 1.03 (N-H) &113.48 & 113.81  & 0.78 &  & 2.28 \\ 
\label{tb:1}
\end{tabular}
\end{table*}

We calculated the work function ($\Phi$) using Equation \ref{eq.EP} as one of the descriptors of the Gibbs free energy for catalytic activity. This electronic property can be measured experimentally, although there is no consensus in the literature on whether a lower or higher work function is more beneficial for HER \cite{chen2024work}. While many studies have suggested that a reduced work function enhances HER activity, others argue that a work function closer to that of noble metals, such as Pd (5.12 eV) or Pt (5.65 eV), leads to improved HER performance \cite{chen2024work, saraf2018pursuit}. Our calculated work function for C$_{36}$N$_{18}$ fell within the value range of noble metals (see Figure \ref{fig:structure_pristine},  Table \ref{tb:1}, and Figure S1 in the SI) with values of 5.27 eV (AA$'$A$''$) and 5.32 eV (ABC). Moreover, it is consistent with experimental results for carbon catalysts with and without nitrogen doping \cite{jia2019identification}.

\begin{figure}[t]
    \centering
    \includegraphics[height=11.8cm]{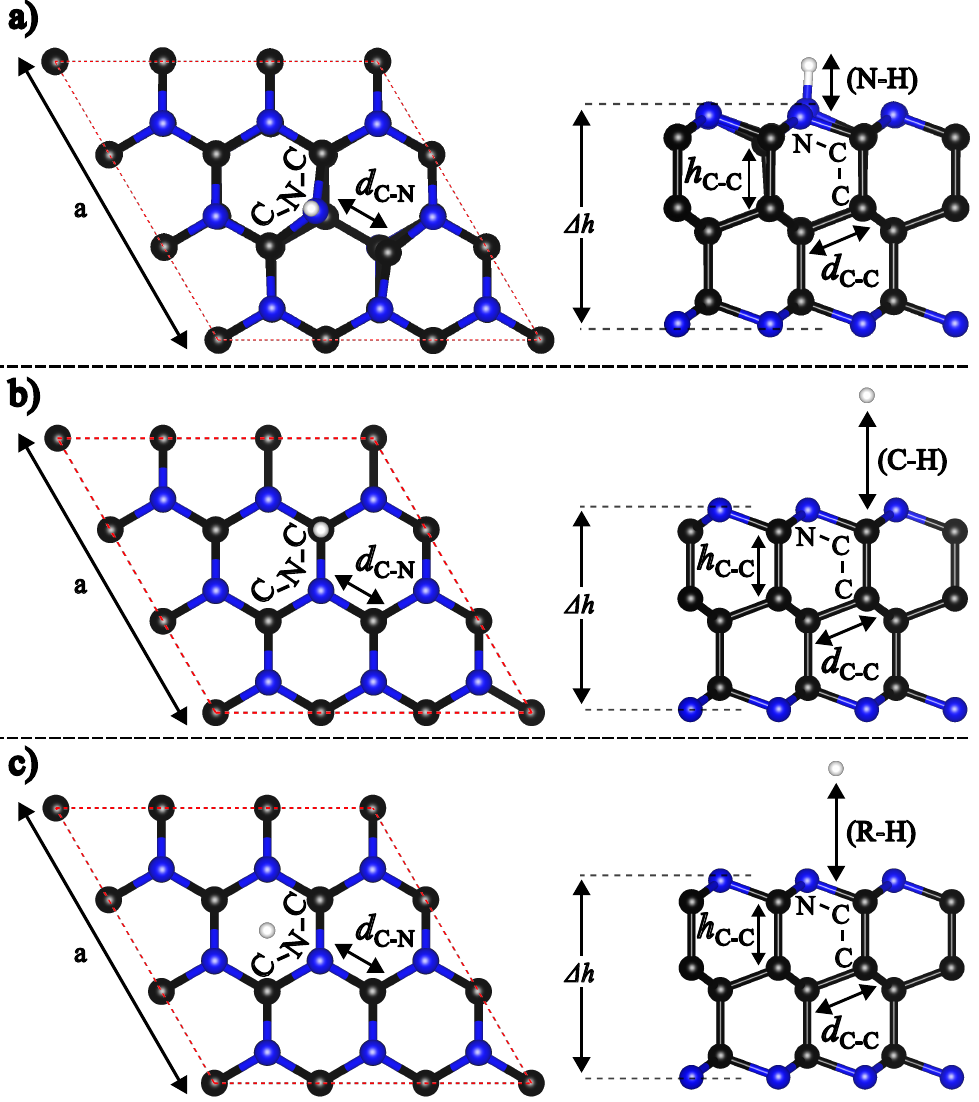}
    \caption{Optimized geometries of hydrogen adsorption sites on pristine AA$'$A$''$-C$_{36}$N$_{18}$, presented in top and side views. The investigated sites include the (a) nitrogen top, (b) carbon top, and (c) hollow sites.  
    Bond distances and structural features upon adsorption are schematically shown and the respective values are in Table \ref{tb:1}. The corresponding details for the ABC stacking configuration are provided in Figure S2 in the SI.}
    \label{fig:structure_pristine_H}
\end{figure}

We next examined whether C$_{36}$N$_{18}$ nanosheets could function as efficient electrocatalysts for HER activity. We investigated the optimal conditions for catalytic performance at three potential sites on C$_{36}$N$_{18}$: the hollow site at the center of a hexagon and the top site directly above each atom, as depicted in Figures \ref{fig:structure_pristine_H} and S2.

After structural optimization, the results show that, in the AA$'$A$''$ stacking configuration, the distances are 3.23 {\AA} for both the carbon top and hollow sites and 1.03 {\AA} for the nitrogen top. The ABC stacking configuration exhibits similar behavior, suggesting that different stacking arrangements impact marginally the hydrogen adsorption, as shown in Table \ref{tb:1}. The interaction between H and C$_{36}$N$_{18}$ indicates physisorption at both the carbon top and hollow sites, as evidenced by the larger distances between H and the surface, with the lattice parameters, bond distances, and angles remaining essentially unchanged with H adsorption. In contrast, the interaction at the nitrogen top presents chemisorption characteristics, where a chemical bond forms between H and N, accompanied by structural changes near the adsorption site (see Figures \ref{fig:structure_pristine_H} and S2, and Table \ref{tb:1}).

The calculated Gibbs free energies (Equation \ref{eq.deltag}) for hydrogen adsorption on pristine C$_{36}$N$_{18}$ are 2.59, 2.59, and 2.23 eV for the AA$'$A$''$ stacking configuration, and 2.59, 2.59, and 2.28 eV for the ABC stacking, at the carbon top, hollow, and nitrogen top sites, respectively. It is well-established that platinum, an ideal catalyst for HER, has a Gibbs free energy close to zero, enabling efficient catalytic activity \cite{chodvadiya2021enhancement}. On the other hand, the significantly positive Gibbs free energy values for pristine C$_{36}$N$_{18}$ indicate that hydrogen adsorption is less favorable, especially on the carbon top and hollow sites. However, stronger hydrogen adsorption was observed at the nitrogen top site, suggesting more promising interactions at that site when compared to the others.

\begin{figure}[t]
    \centering
    \includegraphics[height=10.2cm]{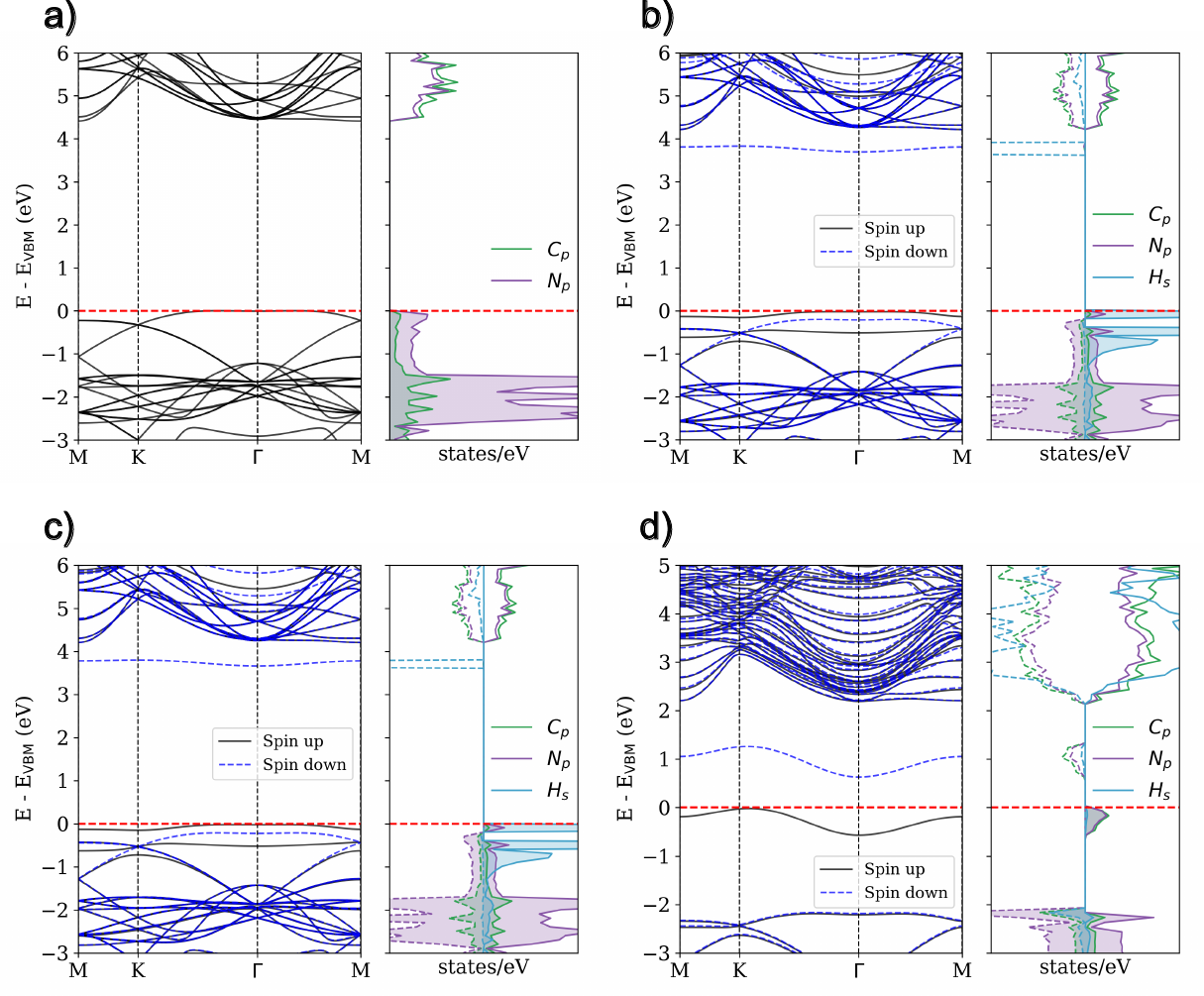}
    \caption{Electronic band structure and PDOS for (a) pristine AA$'$A$''$-C$_{36}$N$_{18}$ and H adsorbed on the (b) carbon top, (c) hollow, and (d) nitrogen top sites. Contributions from carbon, nitrogen, and hydrogen atoms to the valence and conduction bands are highlighted. The corresponding details for the ABC stacking configuration are provided in Figure S3 in the SI.}
    \label{fig:bandstructure_pristine}
\end{figure}

Consequently, we analyzed the electronic band structure and projected density of states (PDOS) for all adsorption sites to gain insight into the hydrogen adsorption mechanism. First, we investigated the pristine C$_{36}$N$_{18}$ systems, as shown in Figures \ref{fig:bandstructure_pristine} and S3. Our calculations revealed a band gap of 4.41 eV for the AA$'$A$''$ stacking and 4.14 eV for the ABC stacking. The PDOS analysis indicates that the nitrogen $p$-orbitals dominate the VBM. In contrast, the CBM consists of a mixture of $s$- and $p$-orbitals from nitrogen and carbon atoms, consistent with previous studies \cite{ipaves2024tuning}.

When hydrogen is adsorbed at the carbon top and hollow sites, the electronic structure is slightly perturbed, resulting in an electron trap state in the spin-down channel due to the H atom (see Figures \ref{fig:bandstructure_pristine} and S3). The PDOS around the Fermi level remains practically unaltered, indicating a weak interaction between hydrogen and the nanosheet, thus confirming the physisorption nature of the interaction at these sites.

In contrast, hydrogen adsorption at the nitrogen top site causes the bond to distort to accommodate nitrogen's $-3$ oxidation state, involving one of the adjacent carbon atoms (see Figures \ref{fig:bandstructure_pristine} and S3). This interaction results in a new band in the spin-up and spin-down channels, with significant contributions from nitrogen and carbon $p$-orbitals. Moreover, all systems exhibit magnetic behavior after hydrogen adsorption, as evidenced by the PDOS asymmetry between the spin-up and spin-down bands. This asymmetry suggests the emergence of magnetism following hydrogen adsorption, particularly at the nitrogen top site.

Table \ref{tb:1} summarizes the calculated structural, electronic, and adsorption parameters related to HER. The similarities in bond distances, Gibbs free energy, electronic band structure, and PDOS imply that stacking 
\begin{figure}[t]
    \centering
    \includegraphics[height=11.9cm]{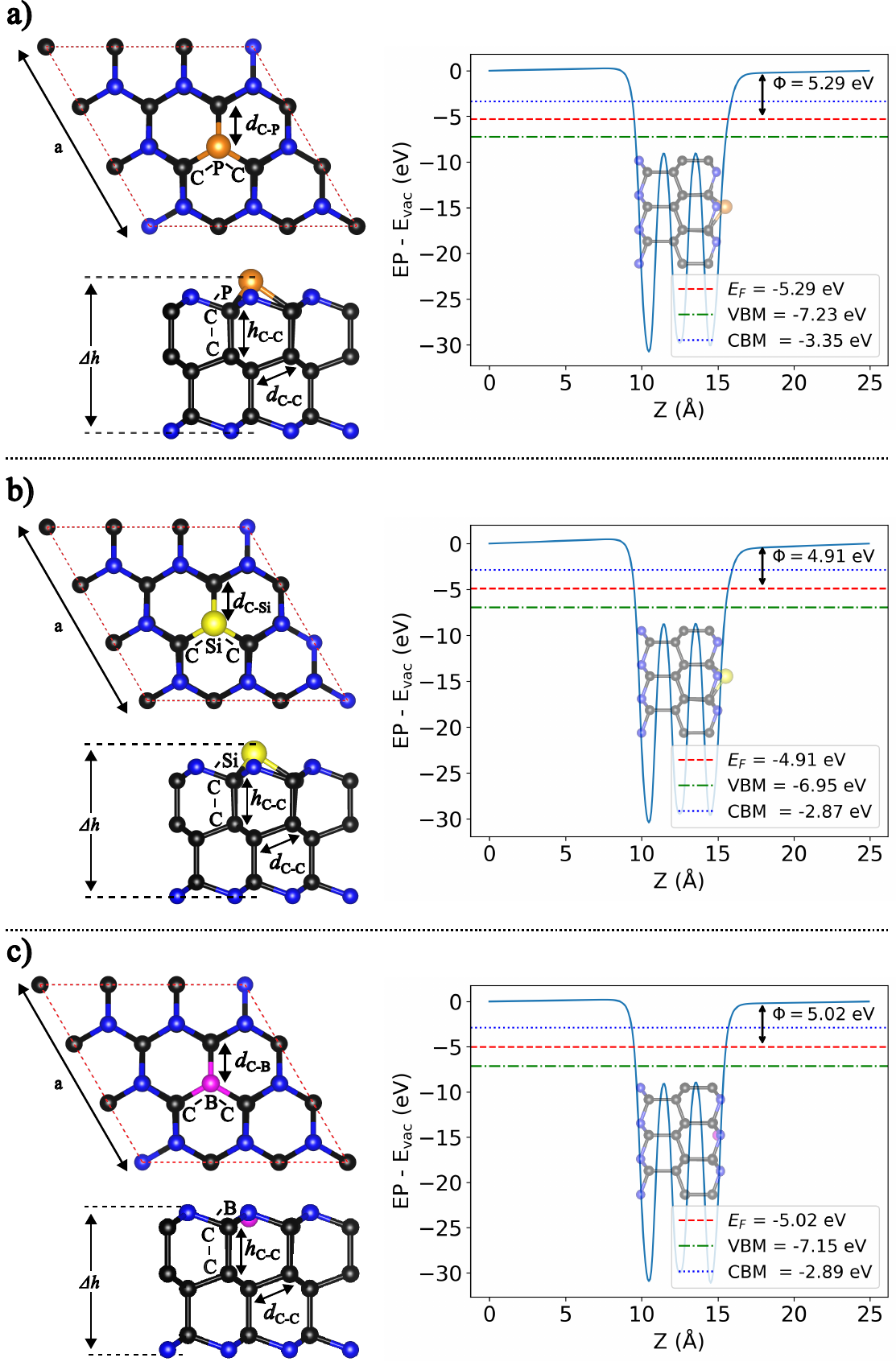}
    \caption{Schematic representation and work function ($\Phi$) of optimized (a) P-doped, (b) Si-doped, and (c) B-doped AA$'$A$''$-C$_{36}$N$_{17}$ nanosheets, presented in top and side views. Structural parameters such as lattice constants, bond lengths, and bond angles are highlighted. The respective values are diplayed in Table \ref{tb:2}. The work function, Fermi energy, VBM, and CBM are referenced relative to the vacuum energy level. The corresponding details for the ABC stacking configuration are provided in Figure S4 in the SI.}
    \label{fig:structure_doped}
\end{figure}

configurations may have a marginal effect on enhancing catalytic activity. Furthermore, based on the Sabatier principle, a high positive Gibbs free energy makes adsorption less favorable, leading to the conclusion that pristine C$_{36}$N$_{18}$ is not well-suited for HER activity. The catalytic efficiency of materials can be enhanced by introducing dopants. Previous studies have demonstrated improved HER activity in g-C$_3$N$_4$ doped with elements such as B, Si, and P \cite{zhu2019catalytic, chodvadiya2021enhancement}. Following the same approach, we investigated the catalytic performance of C$_{36}$N$_{18}$ nanosheets substitutionally doped with either B, Si, or P at the N site at the system's top edge (see Fig. \ref{fig:structure_doped}).

\begin{table*}[t]
\small
\centering
\caption{Structural and electronic properties of doped and hydrogen-adsorbed doped C$_{36}$N$_{17}$: lattice parameter ($a$), intralayer ($d$) and interlayer ($h$) distances, thickness ($\Delta h$), intraplanar (C-X-C) and interlayer (C-C-X) bond angles, electronic band gap ($E_g$), work function ($\Phi$), and Gibbs free energies ($\Delta G$), as labeled in Figures \ref{fig:structure_doped} and \ref{fig:structure_doped_H} (AA$'$A$''$) and S4 and S5 in the SI (ABC). Distances are given in {\AA}, energies in eV, angles in degrees, and X = P, Si, or B.}
\begin{tabular}{lccccccccccc}
 System & $a$&\multicolumn{3}{c}{Bond length} & $\Delta h$ & H Distance & \multicolumn{2}{c}{Angle} & Band Gap & $\Phi$ & $\Delta G$ \\ 
\hline
 & &\multicolumn{3}{c}{} & &  & \multicolumn{2}{c}{} &  &  &  \\ [-3mm]
  AA$'$A$''$    & &$d_{\rm C-C}$ & $h_{\rm C-C}$ & $d_{\rm C-X}$ &  &          & C-C-X & C-X-C &                                  &                 \\
 & &\multicolumn{3}{c}{} & &  & \multicolumn{2}{c}{} &  &  &  \\ [-3mm]
P-doped & 7.32 & 1.50 & 1.59 & 1.82 & 5.27 &  & 122.00 & 90.41  & 3.88 & 5.29 &  \\
H P-doped & 7.32 & 1.50 & 1.59 & 1.82 & 5.27 & 3.16 (P-H) & 122.00 & 90.41  & 3.86 &  & 2.62 \\
Si-doped & 7.33 & 1.51 & 1.58 & 1.82 & 5.21 &  & 119.45 & 92.86  & 4.08 & 4.91 &  \\
H Si-doped & 7.33 & 1.51 & 1.58 & 1.81 & 5.19 & 1.46 (Si-H) & 118.69 & 93.82  & 4.12 &  & -1.51\\
B-doped & 7.31 & 1.50 & 1.60 & 1.51 & 4.74 &  & 99.27 & 116.07  & 4.26 & 5.02 &  \\
H B-doped& 7.31 & 1.50 & 1.61 & 1.58 & 4.87 & 1.20 (B-H) & 111.75 & 105.58 & 3.89 &  & -0.21 \\\\
  ABC  &   &  &  &  &    &         &  & &                 &                 &                 \\
 & &\multicolumn{3}{c}{} & &  & \multicolumn{2}{c}{} &  &  &  \\ [-3mm]
P-doped & 7.35 & 1.51 & 1.57 & 1.83 & 5.20 &  & 121.99 & 90.42 & 3.92 & 5.18 &  \\
H P-doped & 7.35 & 1.51 & 1.57 & 1.83 & 5.20 & 3.14 (P-H) & 121.99 & 90.42 & 3.89 &  & 2.58 \\
Si-doped & 7.37 & 1.51 & 1.57 & 1.82 & 5.15 &  & 119.35 & 92.99  & 4.00 & 4.99 &  \\
H Si-doped & 7.37 & 1.51 & 1.56 & 1.81 & 5.12 & 1.46 (Si-H) & 118.70 & 93.87  & 3.87 &  & -1.50\\
B-doped & 7.34 & 1.50 & 1.61 & 1.51 & 4.65 &  & 99.65 & 115.87  & 4.26 & 4.94 &  \\
H B-doped& 7.34 & 1.50 & 1.58 & 1.58 & 4.78 & 1.20 (B-H) & 111.81 & 105.64  & 3.78 &  & -0.26  \\ 
\label{tb:2}
\end{tabular}
\end{table*}

Upon substituting nitrogen with dopants, optimization revealed that lattice parameters ($a$) and bond distances ($d_{\rm C-C}$ and $h_{\rm C-C}$) had only minor variations. In contrast, $d_{\rm C-X}$ distances and C-C-X and C-X-C bond angles underwent significant changes, indicating that doping induced only localized structural perturbations (see Figures \ref{fig:structure_doped} and S4, and Table \ref{tb:2}). The effects of P and Si substitutions on local structural parameters were similar, adjusting $d_{\rm C-X}$ distances and altering the C-C-X and C-X-C bond angles to nearly 90° for C-X-C, reflecting an out-of-plane strain and an increase in $\Delta h$. Here, $\Delta h$ refers to the vertical distance between the dopant and the nitrogen atom on the opposite edge, with P and Si positioned further from the surface than N. However, B substitution produced a more planar C-X-C angle close to 120°, characteristic of $sp^2$ hybridization, resulting in an almost flat local structure (see Table \ref{tb:2}, Figures \ref{fig:structure_doped} and S4).
\begin{figure}[b!]
    \centering
    \includegraphics[height=11.2cm]{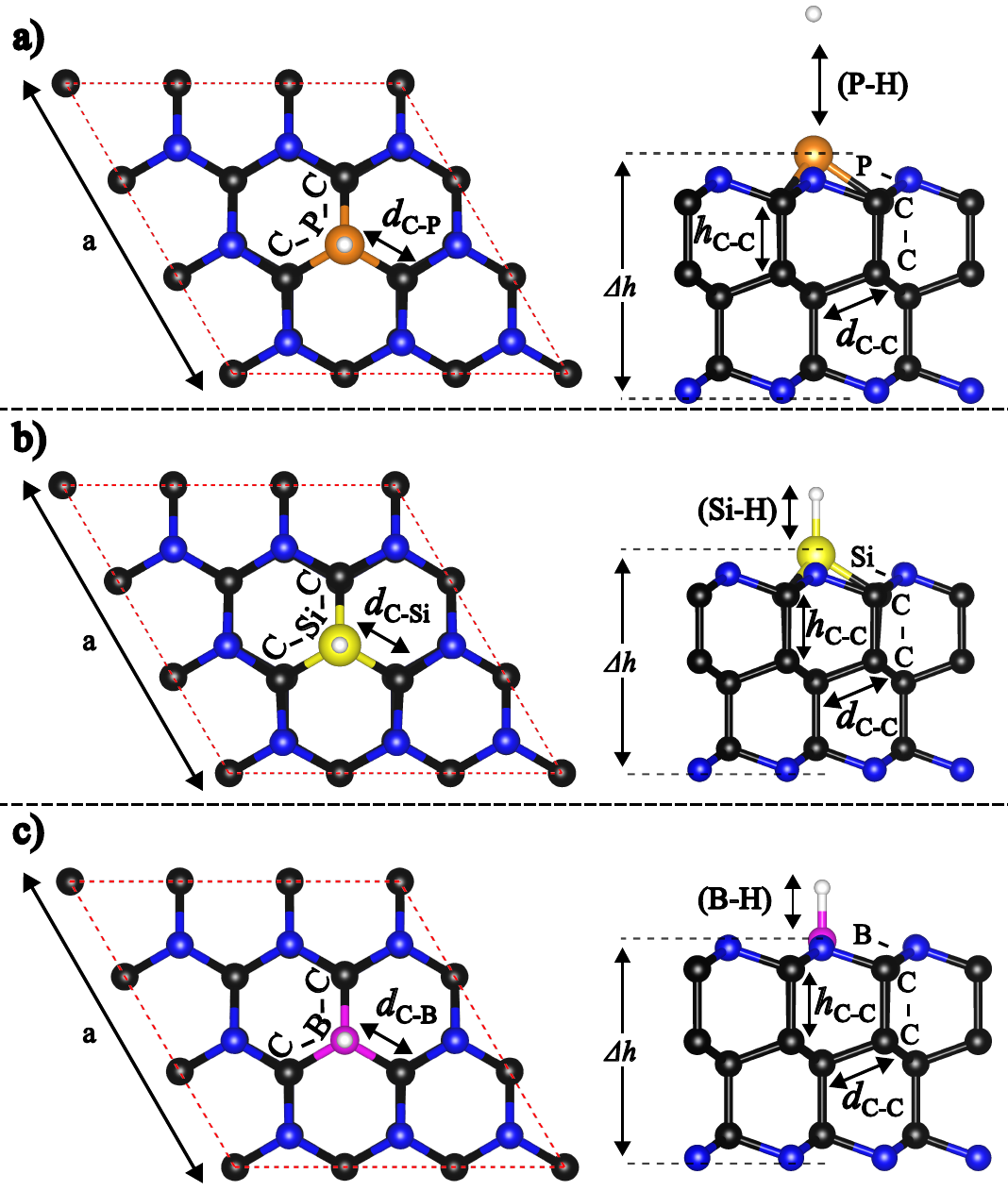}
    \caption{Optimized geometries of hydrogen adsorption sites on (a) P-doped, (b) Si-doped, and (c) B-doped AA$'$A$''$-C$_{36}$N$_{17}$, presented in top and side views. Changes in bond distances and structural features upon adsorption are shown. The respective values are in Table \ref{tb:2}. The corresponding details for the ABC stacking configuration are provided in Figure S5  in the SI.}
    \label{fig:structure_doped_H}
\end{figure}

The effects of P and Si substitutions on local structural parameters were similar, adjusting $d_{\rm C-X}$ distances and altering the C-C-X and C-X-C bond angles to nearly 90° for C-X-C, reflecting an out-of-plane strain and an increase in $\Delta h$. Here, $\Delta h$ refers to the vertical distance between the dopant and the nitrogen atom on the opposite edge, with P and Si positioned further from the surface than N. However, B substitution produced a more planar C-X-C angle close to 120°, characteristic of $sp^2$ hybridization, resulting in an almost flat local structure (see Table \ref{tb:2}, Figures \ref{fig:structure_doped} and S4).

Following the approach used for the pristine structure, we calculated $\Phi$ (Equation \ref{eq.EP}) for all doped configurations. The calculated work functions for the doped C$_{36}$N$_{17}$ systems showed variations influenced by the type of dopant (see Table \ref{tb:2}). For the P-doped system, the $\Phi$ value changed slightly, remaining within the range of noble metals, similar to the pristine case. On the other hand, the Si- and B-doped systems exhibited a reduction in $\Phi$, with values around 5.00 eV. This reduction in $\Phi$ could contribute to high catalytic activity \cite{jia2019identification}.

We then adsorbed a single H atom on top of each dopant. In the final configuration, the H atom induced minor structural changes in the P- and Si-doped systems, while the B-doped case experienced a more notable shift, with bond angles transitioning from nearly $sp^2$ to almost $sp^3$ hybridization (see Table \ref{tb:2} and Figures \ref{fig:structure_doped_H} and S5). Notably, the H atom did not bond with the P atom, maintaining a P–H distance of approximately 3.15 Å, displaying physisorption behavior similar to that observed in the pristine case when H was adsorbed at the carbon top and hollow sites. Nevertheless, the interaction for the Si and B top sites resulted in a chemical bond between H and the dopant atoms, with H-dopant distances of 1.46 Å and 1.20 Å, respectively, indicating chemisorption. As for the pristine case, both stackings exhibited similar behavior, suggesting that different arrangements still do not significantly influence hydrogen adsorption in the doped systems.

The calculated Gibbs free energies (Equation \ref{eq.deltag}) for H adsorption on the doped structures are summarized in Table \ref{tb:2}. Notably, the B-doped system shows a marked improvement compared to pristine C$_{36}$N$_{18}$, with Gibbs free energy values approaching zero, indicative of efficient catalytic activity. Nevertheless, the P-doped system displays significantly positive Gibbs free energy values, while the Si-doped system exhibits highly negative values. This underscores the potential for enhanced catalytic activity through doping. To investigate the underlying reasons for these differences, we analyzed the electronic structure and PDOS of the doped systems before and after H adsorption.

\begin{figure}[hbt]
    \centering
    \includegraphics[height=13cm]{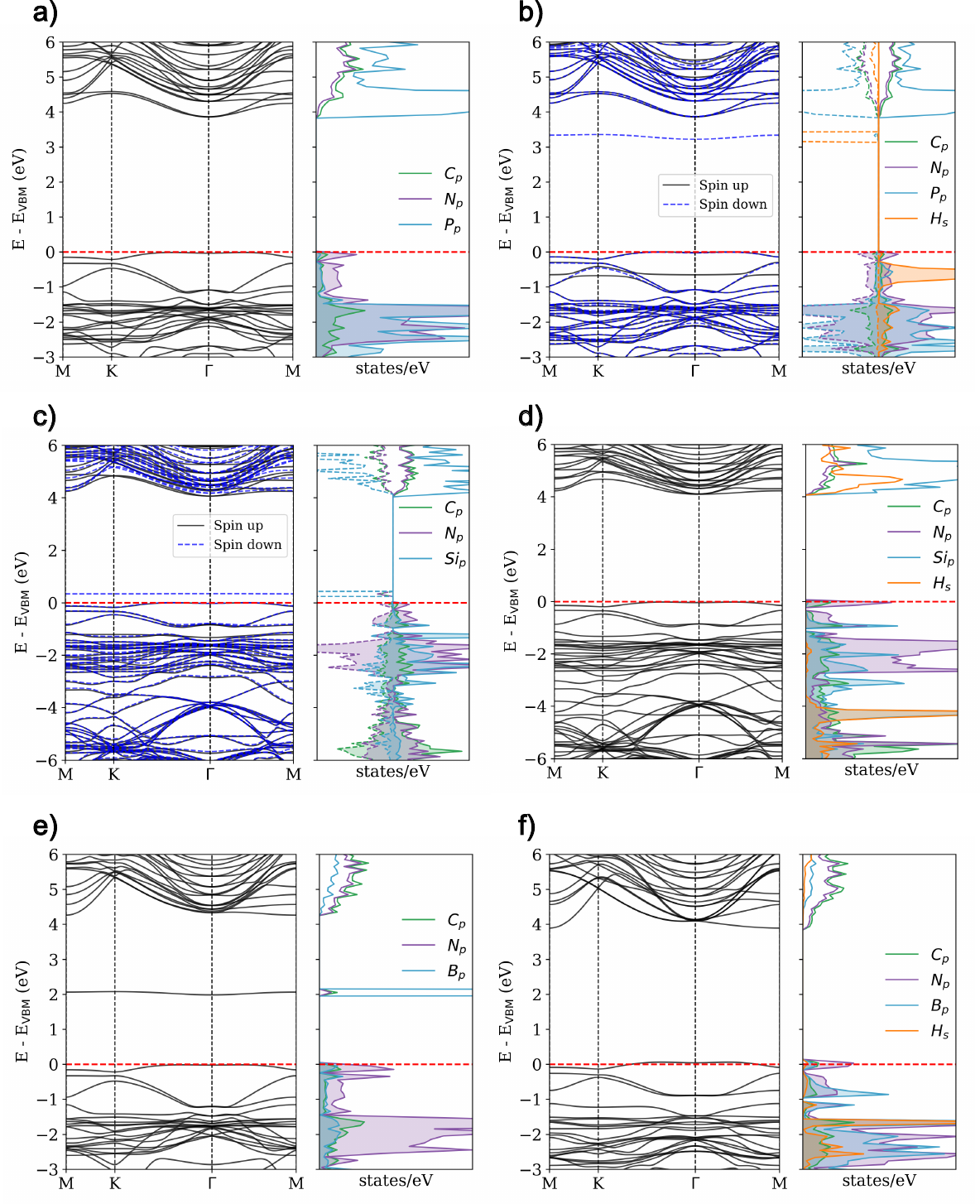}
    \caption{Electronic band structure and PDOS for (a) P-doped, (b) H adsorbed on P-doped, (c) Si-doped, (d) H adsorbed on Si-doped, and (e) B-doped, (f) H adsorbed on B-doped AA$'$A$''$-C$_{36}$N$_{17}$. Contributions from carbon, nitrogen, phosphorus, silicon, boron, and hydrogen atoms to the valence and conduction bands are highlighted. Details for the ABC stacking configuration can be found in Figure S6  in the SI.}
    \label{fig:bandstructure_doped}
\end{figure}

Figures \ref{fig:bandstructure_doped} and S6 present the band structures and PDOS before and after H adsorption, while Table \ref{tb:2} shows key properties. The P-doped case shows minimal changes in the band structure and PDOS character since P and N are isovalent. However, the PDOS for the P-doped system reveals contributions from $p$-orbitals due to P-C bonding, with a reduced band gap to approximately 3.90 eV. In the Si-doped case, the replacement of N by Si, which has one less valence electron than N, introduces notable changes. The main effect is the emergence of magnetism, as evidenced by the asymmetry between the PDOS for spin-up and spin-down electrons. Additionally, a localized state appears in the spin-down channel, primarily due to the $p$-orbital of the Si atom. The band gap is reduced to around 4.00 eV if the localized state is not considered as effective narrowing of the gap. The B-doped system does not exhibit magnetism. With three valence electrons, B forms a localized state in the middle of the gap, primarily due to its $p$-orbitals. This results in a gap reduction to approximately 4.26 eV, similar to the Si case, where we consider this localized state not as an effective gap narrowing.

After H adsorption, the band structure and PDOS changed, as shown in Figures \ref{fig:bandstructure_doped} and S6. For the P-doped system, physisorption occurs, as evidenced by slight changes in the band structure and PDOS compared to the pre-adsorption state, along with two localized bands associated with the s-orbital of the H atom. Specifically, an s-orbital state in the spin-up channel is present within the valence band, and an s-orbital trap of H is in the spin-down channel close to the conduction band. This configuration introduces magnetism due to the unpaired electron of the H atom.

The Si- and B-doped systems, which initially exhibited trap states before H adsorption, undergo H chemisorption, with H atoms binding within these traps and effectively eliminating magnetism in both cases. For the Si-doped system, the band gap remains nearly unchanged, while a strong bond forms between H and Si. This is evidenced by the deeply bound s-orbitals of H observed in the PDOS, which likely contributes to the highly negative $\Delta G$. In contrast, in the B-doped system, a partially filled band emerges at the top of the valence band, as revealed by the PDOS, which highlights hybridization between the H atom and nearby atoms. Similar partially filled bands have been observed in $\gamma$-graphyne after doping with B atoms, indicating a semiconductor-to-metal transition \cite{ruiz2017dft}. This possible transition could explain the $\Delta G$ lower value  in the B-doped system compared to that of the Si-doped case, emphasizing the role of dopant selection in optimizing catalytic performance.

\begin{figure}[hbt]
    \centering
    \includegraphics[height=9.4cm]{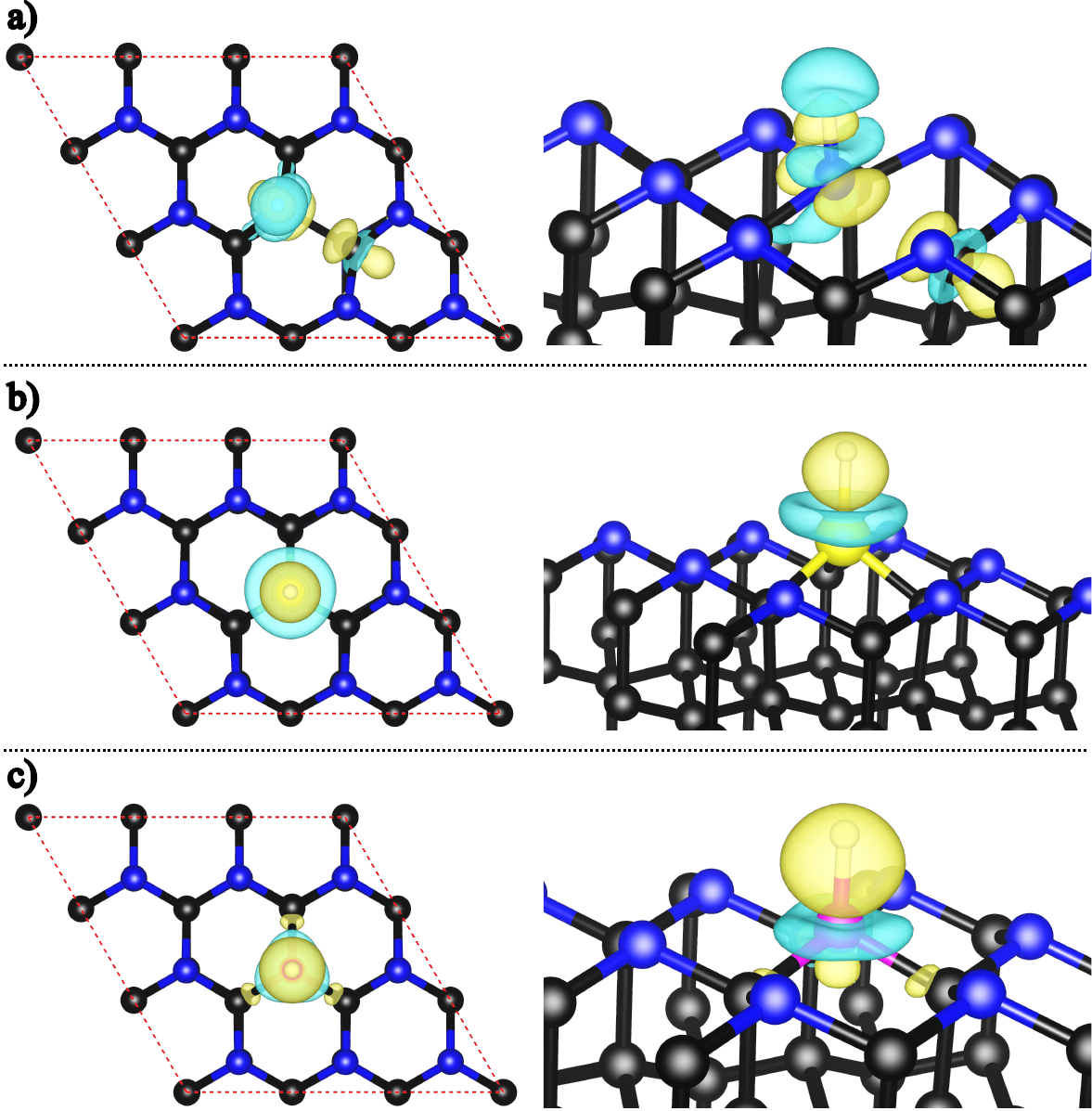}
    \caption{Charge density differences of hydrogen adsorption sites on (a) N-top of pristine (isosurface value 0.01 e/Bohr$^3$), (b) Si-doped (isosurface value 0.0055 e/Bohr$^3$), and (c) B-doped (isosurface value 0.012 e/Bohr$^3$) AA$'$A$''$-C$_{36}$N$_{17}$, shown in top and side views. Charge accumulation (yellow) and depletion (cyan) regions are highlighted. Details for the ABC stacking configuration can be found in Figure S7 in the Supporting Information.}
    \label{fig:charge}
\end{figure}

To investigate the chemisorption interactions between H and both pristine and doped systems, focusing on cases where H binds to N, Si, or B atoms, we computed the charge density difference (see Equation \ref{eq:charge_diff} and Figures \ref{fig:charge} and S7) and Bader charge analyses. In the pristine C$_{36}$N$_{18}$, charge accumulation, highlighted in yellow, appears between the H and N atoms and near a C atom that distorts upon bonding to H. The Bader charge analysis reveals a charge transfer of 0.54 $|e|$ from the H atom to C$_{36}$N$_{18}$, which is an indication of a polar covalent bond. For the Si- and B-doped systems, the charge distribution varies notably: in the Si-doped structure, the charge concentration centers around the H atom, while in the B-doped case, accumulation occurs beneath the B atom and between the B and neighboring C atoms. Here, Bader analysis indicates a charge transfer of 0.51 $|e|$ from the H atom to the Si-doped C$_{36}$N$_{17}$ and 0.54 $|e|$ to the B-doped C$_{36}$N$_{17}$, reflecting a shift in the interaction as compared to the pristine structure. These results suggest that Si and B dopants alter the electronic environment, enhancing the interaction with hydrogen and potentially boosting the catalytic efficiency.

Table \ref{tb:2} summarizes the structural, electronic, and adsorption parameters related to HER performance for doped systems. Stacking configurations show minimal impact on catalytic enhancement, as evidenced by the similarity in bond distances, Gibbs free energy, electronic band structure, and PDOS. Based on the Sabatier principle, the B-doped system, with a $\Delta G$ around -0.2 eV, appears well-suited for HER activity.
Although the B-doped system showed the most favorable $\Delta G$, the applicability of the work function as a reliable descriptor remains uncertain. As discussed previously, there is no consensus on whether a lower or higher $\Phi$ is more beneficial for HER. A higher work function appears favorable in the B-doped case, where a partially filled band is present after H adsorption.

Finally, achieving effective electrocatalysis requires the system to exhibit metallic behavior, while an ideal photocatalytic bandgap should be greater than 1.23 eV and must satisfy the band edge (VBM and CBM) alignment requirements \cite{rahman2020metal}. Accordingly, B-doped C$_4$N$_2$ could be an efficient cocatalyst when paired with a metallic material to enhance overall catalytic performance. In this configuration, the metallic component would supply electrons and enable efficient charge transfer, while the B-doped C$_4$N$_2$ nanosheet would provide active sites for H adsorption. This design strategy aligns with established approaches in photocatalytic and electrocatalytic systems, where cocatalysts optimize surface reactions, while the primary catalyst ensures effective electron transport and catalytic efficiency \cite{rahman2020metal, liu2022review}.

\section{Conclusions}

In conclusion, this study investigated the structural, electronic, and catalytic properties of pristine and doped C$_4$N$_2$ nanosheets as potential electrocatalysts for the HER. Our results indicate that pristine C$_{36}$N$_{18}$ nanosheets exhibit limited HER activity due to their high positive Gibbs free energy. Doping with P maintains this limited activity, as the Gibbs free energy remains highly positive. In contrast, Si doping leads to highly negative Gibbs free energies, making hydrogen adsorption too strong for effective HER catalysis. The most promising results were observed for B-doped C$_{36}$N$_{17}$ nanosheets, which exhibit a Gibbs free energy close to zero, indicating optimal hydrogen adsorption for efficient HER. Charge density and Bader charge analyses reveal substantial changes in the electronic environment upon doping. The modifications caused by B-doping significantly alter the C$_{36}$N$_{18}$  electronic structure in a way that enhances catalytic performance. Although different stacking configurations (AA$'$A$''$ and ABC) have minimal impact on HER activity, B-doping emerges as a key strategy to optimize hydrogen adsorption. These findings suggest that B-doped C$_{36}$N$_{17}$ nanosheets could serve as efficient cocatalysts when combined with metallic materials, offering a promising approach to enhancing overall catalytic efficiency in electrocatalytic and photocatalytic systems.

\section*{Conflicts of interest}
There are no conflicts to declare.

\section*{Acknowledgements}
P.A.S.A., J.M.A., and B.I. thank CNPq - INCT (National Institute of Science and Technology on Materials Informatics, grant 371610/2023-0), the UFABC Multiuser Computational Center (CCM-UFABC), the National Laboratory for Scientific Computing (LNCC/MCTI, Brazil) and Centro Nacional de Processamento de Alto Desempenho em São Paulo (CENAPAD-SP, Brazil) for providing HPC resources used in this work. J.F.J thanks CNPq (grant 302800/2022-0) L.V.C.A.  acknowledge funding from CNPq (Project 314884/2021-1) and support from FAPESP (Project 2022/10095-8).

\bibliography{database}

\end{document}